\documentclass[twoside]{article}
\usepackage{fleqn,espcrc2,epsfig,axodraw}
\def\prd#1{{\em Phys.~Rev.}~{\bf D#1}\ }
\def\prl#1{{\em Phys.~Rev.~Lett.}~{\bf #1}\ }

\def\np#1{{\em Nucl.~Phys.}~{\bf B#1}\ }
\def\deg{\ifmmode{^{\circ}}\else ${^{\circ}}$\fi}
\def\ni#1{\noindent$(#1)\quad$}

\def\lsim{\,\raisebox{-0.13cm}{$\stackrel{\textstyle<}{\textstyle\sim}$}\,}
\def\bi{\begin{itemize}}
\def\ei{\end{itemize}}
\def\ed{\end{document}}
\def\be{\begin{equation}}
\def\ee{\end{equation}}
\def\beas{\begin{eqnarray*}}
\def\eeas{\end{eqnarray*}}
\def\vev#1{\left<{#1}\right>}
\def\bm#1{\ifmmode{\mbox{\boldmath $#1$}}\else {\boldmath $#1$}\fi}
\def\mpl{\ifmmode{M_{\rm Pl}}\else ${M_{\rm Pl}}$\fi}
\def\mgut{\ifmmode{M_{\rm GUT}}\else ${M_{\rm GUT}}$\fi}

\def\sug{SUSY-GUTs}
\def\bst{{\bf 16} }
\def\bten{{\bf 10}}
\def\bfv{{\bf 45}}
\def\bots{\overline{\bf 126}}

\def\eb{\end{thebibliography}}

\def\ra{\rightarrow}
\def\mueg{\ifmmode{\mu\ra e\gamma} \else $\mu\ra e\gamma$\fi}
\def\taumg{\ifmmode{\tau\ra \mu\gamma}\else $\tau\ra \mu\gamma$\fi}
\def\taueg{\ifmmode {\tau\ra e\gamma}\else $\tau\ra e\gamma$\fi}

\def\vs#1{\vspace*{#1cm}}

\newcommand{\AmS}{{\protect\the\textfont2
  A\kern-.1667em\lower.5ex\hbox{M}\kern-.125emS}}

\title{How Georgi-Jarlskog and  SUSY-SO(10) imply a measurable rate for 
$\mueg$}

\author{Haim Goldberg \thanks{Research supported in part by 
Grant No PHY-9411546 from the 
        National Science Foundation.} and Mario E. G\'{o}mez  \\ [2.5mm]
   Department of Physics, Northeastern University\\ 
       Boston, MA 02115}
       
\begin{document}

\begin{abstract}
 Slepton mass matrices have been analyzed in an SO(10) SUSY-GUT, with
 soft-breaking terms generated at Planck scale. Higher dimensional operators
 consistent with 4-d string constructions are used
 in order to generate a Georgi-Jarlskog (G-J) Yukawa texture at \mgut. Radiative
 corrections between \mpl\ and \mgut\ generate a substantial 
 non-universality,  in
 the $\tilde\mu-\tilde e$ sector. This non-universality originates in the
 flavor dependence of the Higgs assignments required for the G-J texture, and
 is 
 {\em unrelated to the
 large top Yukawa}. The resulting branching ratio for $\mueg$ could make
 this process observable  for large sectors of the MSSM  parameter space, with a
 factor of 10 improvement in statistics.
\end{abstract}
\maketitle

\section{INTRODUCTION} Supersymmetric Grand Unified Theories (\sug), in
combination with  phenomenologically acceptable {\em ansatze} about Yukawa
structures at GUT, can provide some interesting predictions for low energy CKM
parameters. A second (logically disconnected) feature of \sug\ emerges when
soft terms are generated at the Planck scale through the coupling to a hidden
sector via  supergravity: even if these soft-breaking terms are universal at
\mpl,  radiative corrections between \mpl\ and \mgut\ may induce enough flavor
dependence in the soft-breaking structure at GUT to be reflected in
important low-energy flavor violation \cite{Hal,Gab}. 

Heretofore, these corrections
have  been almost exclusively reflective of the large top quark
coupling -- this has differentiated the generations. This talk will
contain a synopsis of some recent work \cite{Gom}
which couples the two features in the
last paragraph in a fashion leading to a new source flavor violation in the
lepton sector. The argument, briefly, is that phenomenologically 
acceptable fermion textures at
GUT (such as G-J) are compatible with  specific
flavor dependence in the Higgs assignments above
\mgut; 
these in turn generate non-universality in the soft structure at GUT, which has
phenomenologically significant consequences for the process
$\mueg.$ These consequences are entirely disconnected from 3rd generation
dynamics, and are the core subject of this talk. Full  details are given in the
published work \cite{Gom}. 

\section{THE MODEL}
The fermion mass matrices at GUT which incorporate the G-J texture are 
\begin{eqnarray}
 h^u&=&\left(\begin{array}{ccc} 0&C&0 \\ C&0&B\\ 0&B&A
\end{array}\right),\ \ 
h^d=\left(\begin{array}{ccc} 0&F&0 \\ F&E&0\\ 0&0&D
\end{array}\right),\nonumber\\
&&h^e=\left(\begin{array}{ccc} 0&F&0 \\ F&-3 E&0\\ 0&0&D
\end{array}\right)\ .
\end{eqnarray}
The 2-2 entries in $h^d,\ h^e$ allow for a reasonable low energy value of
$m_s/m_{\mu}.$ In SO(10) \sug, a $\bst_2\bst_2\bots$ term in the
superpotential will provide for the G-J texture. However, a $\bots$ is
disallowed in 4-d string constructions\cite{Cha}, and leads to
a rapid breakdown of perturbation theory above GUT. Hence, we are led to
use  a Yukawa structure containing composite operators which will
substitute for the $\bots.$ These operators can be pictured as the result of
integrating out some set of heavy superfields at a  scale $M\ \lsim\ \mpl.$ For
definiteness, the
model used will be the one proposed by Babu and Mohapatra\cite{Bab}with a
Yukawa superpotential
\begin{eqnarray}
W_{yuk}&=&M^{-1}\bst_3\bst_3(Y_{33}\bten_dS_1+h_{33}\bten_uS_2)\nonumber\\
&+&M^{-1}\bst_2\bst_3\ (h_{23}\bten_u S_3)\nonumber\\
&+&M^{-2}\bst_2\bst_2\ (Y_{22}\bfv_1\bfv_2\bten_d)\nonumber\\
&+&M^{-3}\bst_1\bst_2(Y_{12}\bten_d S_2^3+h_{12}\bten_u S_1^3)\nonumber\\
\end{eqnarray}
There are discrete symmetries which prescribe the couplings shown (including
those of the gauge singlets $S_{1,2,3}$), and both these and the SO(10)
gauge symmetry are broken at
\mgut. All vevs (except for
$\vev{S_2}\sim M$) are of order
\mgut, which sets up the proper Yukawa hierarchy. (All the coupling constants
are of $O(1).)$ The vev structure of the
${\bf 45}$'s ($\vev{\bfv_1}||(B-L),\ \vev{\bfv_2}||\ T_{3R}),$ leads to
an effective
$\bots$ in the product $\vev{\bfv_1\bfv_2\bten_d}. $ As an example, the Yukawa
coupling $B$ in Eq.~(1) is given by $h_{23}\vev{S_3}/M.$

\section{SOFT-BREAKING TERMS} In this work we assume universal soft scalar
masses $m_0^2$ and gaugino masses $M_{1/2}$ at \mpl, and multinomial scalar
terms
proportional to terms in 
$W_{yuk}$ between
$M\sim \mpl$ and \mgut:
\be -{\cal L}_{soft}=M^{-1}\overline{Y}_{33}\bst_3\bst_3\bten_d S_1+\ldots\
,\ee
where now the fields stand for their scalar components.
After symmetry breaking,
this provides trilinear terms 
$\xi_{ij}L_iH_dE_j$\ , where 
\[(\xi_{33})_{\ \rm GUT}=M^{-1}\ \overline{Y}_{33}^{\rm GUT}\vev{S_1}\ ,
\mbox{\em etc.}\]

\section{NON-UNIVERSALITY AT \mgut.} There are two sources of 
non-universality at
\mgut:
\smallskip
 
\ni{1}There are flavor-dependent corrections $\delta m^2_{16}$ to the
$\bst_i^*\bst_j$ scalar mass matrix. The off-diagonal elements are 
tied directly 
to the large $t$-quark coupling, and originate in 
the integration of diagrams such as Fig. 1
between \mpl\ and \mgut.\vs{.4}

\begin{center}
\begin{picture}(150,100)(0,0)
\DashCArc(75,75)(25,0,360){4}\Text(85,100)[lb]{${\bf 16}_{3}$,\ ${\bf 10}
_{u}$}
\Line(72,97)(78,103)
\Line(72,103)(78,97)
\DashArrowLine(0,50)(75,50){4}\Text(30,60)[rc]{${\bf 16}_{3}$}
\DashArrowLine(75,50)(145,50){4}\Text(145,60)[lc]{${\bf 16}_{2}$}
\DashArrowLine(75,50)(125,0){4}\Text(115,30)[rc]{$S_3$}
\DashArrowLine(25,0)(75,50){4}\Text(35,30)[lc]{$S_2$}
\end{picture}
\end{center}\smallskip

\noindent Figure 1. One loop contribution to the 3-2 entry of 
$\delta m_{16}^2\ .$\vs{.4}

The result of such integrations (with details given in
Ref.~\cite{Gom}) gives a contribution to the slepton mass matrix at GUT
\begin{eqnarray}
\delta m_{16}^2&=&-10\ \frac{m_{16}^2}{8 \pi^2}
\ \log\left(\frac{\mpl}{\mgut}\right)\nonumber\\
& &\qquad\cdot\quad\left(\begin{array}{ccc} 
Y_{12}^2 & 0 & BC \\
0 & Y_{12}^2 & 2 BA\\ 
BC & 2 A B  & 4 A^2
\end{array} \right)\end{eqnarray}
Note that the third generation receives a large contribution $(\delta
m_{33}^2)$ from the top Yukawa A. (It turns out that  $A\approx 3.$)
This leads to a significant (indeed, nonperturbative) splitting of
$m_{\tilde{\tau}}^2$ from
$m_{\tilde{\mu}}^2\simeq m_{\tilde{e}}^2$. (This possibility was noted in Ref.
\cite{Bar}). In the present model, flavor violations in the
$\mu-e$ sector will not involve the $\tilde{\tau}$ or its mass. Flavor
violations in
$\tau-\mu$ or
$\tau-e$ processes will involve loops with $\tilde{\tau},$ and this mass
splitting will be simply parameterized when these processes are discussed.
\smallskip

\ni{2}Because the SO(10) content of the Higgs structure contributing to the 2-2
matrix element of the superpotential is different from that contributing to 
 the other
entries, there will be flavor dependence induced in the $\mu-e$ sector of the
effecive trilinear  matrix $\xi_{ij}$ at GUT. The radiative corrections
pertinent to this part of the soft-breaking are shown in Fig. 2.\vs{.4} 
\begin{center}
\begin{picture}(100,100)(0,0)
\DashArrowLine(15,85)(24,76){2}\Text(15,85)[rc]{${\bf 16_2}$}
\ArrowLine(24,76)(50,50)
\DashArrowLine(15,15)(24,24){2}\Text(15,15)[rc]{${\bf 16_2}$}
\ArrowLine(24,24)(50,50)
\DashArrowLine(85,85)(50,50){4}\Text(85,85)[lc]{${\bf 45_1}$}
\DashArrowLine(100,50)(50,50){4}\Text(100,50)[lc]{${\bf 45_2}$}
\DashArrowLine(85,15)(50,50){4}\Text(85,15)[lc]{${\bf 10_d}$}
\ArrowLine(24,50)(24,24)
\ArrowLine(24,50)(24,76)
\Line(21,53)(27,47)
\Line(21,47)(27,53)
\end{picture}
\end{center}
\smallskip

\noindent Figure 2. One-loop contribution to the effective trilinear $(\xi)$ at
GUT not in common with one-loop renormalization of effective Yukawa. Not shown
are additional figures with gaugino loops for all pairs of legs.\vs{.4}
 
Together with
the tree contribution, they lead to the following expression for the
soft trilinear matrix $\xi_e$  at \mgut\ (after SO(10) breaking):
\begin{eqnarray}(\xi_{ab}^e)_{G}&=&m_0 A_{0}
\left(\begin{array}{ccc}
 0 & F & 0 \\
 F & -3 E & 0\\
0 & 0 & D
\end{array}\right)\nonumber\\
&-&63\frac{\alpha_{\rm GUT}}{4\pi}\ M_{1/2}\
  \log\left(\frac{\mpl}{\mgut}\right)\nonumber\\
& &\cdot\quad\left( \begin{array}{ccc} 0&F&0\\ F&(127/63)(-3E)&0\\ 0&0&D
\end{array}\right)
\end{eqnarray}

\bi\item We now observe that neither $\delta m_{16}^2$ (Eq.~(4))
nor $\delta\xi$ (second term in Eq.~(5) is proportional
to the Yukawa matrix, so that these will prove a source of lepton flavor
violation.\ei
\section{THE INSERTIONS}On going to a lepton flavor-diagonal basis,
one obtains from  Eqs.~(4) and (5) the slepton flavor 
changing insertions (ignoring
$\tau-e$ mixing)
\begin{eqnarray}
\Delta_{\tau\mu}^{LL}&=&\Delta_{\tau\mu}^{RR}\simeq
-20\ BA\ \frac{m_{16}^2}{8 \pi^2}\ \log\left(\frac{\mpl}{\mgut}\right)
\nonumber\\
\Delta_{\mu e}^{LR}&=&\Delta_{\mu e}^{RL}\simeq 
64\ \frac{\alpha_{\rm GUT}}{4\pi}\sqrt{m_e m_{\mu}}\nonumber\\
& &\hspace*{1cm}\cdot\ \ M_{1/2}\
  \log\left(\frac{\mpl}{\mgut}\right)\ \ .\end{eqnarray}
As advertised, the $\mu-e$ insertion is independent of all 3rd generation
physics. Moreover, it is {\em independent} of the MSSM $\mu$ parameter. The fact
that it is non-zero and large {\em both} originate in the Georgi-Jarlskog
constraint.\footnote{The parametric form of the $\mu\ra e$ transition described
in Ref.~\cite{Bar} is the same as in Eq.~(6); however, the model used by these authors
does not incorporate a realistic Yukawa texture.}

\section{PARAMETERS}
The Feynman diagrams of Fig.~(3) are then calculated, with the  $\Delta$'s of
Eq.~(6)  as insertions. The Yukawa parameters $(A-F)$ at \mgut\ are obtained
by running to low energies and fitting to fermion masses and mixing. 
The rest of the parameters necessary
(chargino and  neutralino masses and mixing angle, and slepton  masses) are
obtained as functions of a two parameter space: $M_G$, the gaugino mass at
\mgut, and $m_{\tilde{\mu}_L},$ the smuon (and selectron) mass at low
energies. We have fixed $\tan\beta=3,\;A_0=1,\ \mbox{and}\ \mu>0.$ 
\vs{2}

\begin{center}
\begin{picture}(150,150)(0,0)
\ArrowLine(0,130)(20,130)\Text(10,138)[cc]{{\scriptsize $\tau_L$}}
\ArrowLine(40,130)(20,130)\Text(30,138)[cc]{{\scriptsize $\tau_L^c$}}
\ArrowLine(150,130)(110,130)\Text(140,138)[lc]{{\scriptsize $\mu_L^c$}}
\Photon(115.4,179.42)(97,144.8){2}{7}\Text(122,185)[lt]{{\scriptsize $\gamma$}}
\DashCArc(75,109.8)(40.4,60,90){4}\Text(95,150)[rb]{{\scriptsize 
 $\tilde{\mu}_{R}$}}
\DashCArc(75,109.8)(40.4,30,60){4}
\DashCArc(75,109.8)(40.4,90,150){4}\Text(60,150)[rb]{{\scriptsize
$\tilde{\tau}_{R}$}}
\Line(72,153.2)(78,147.2)
\Line(72,147.2)(78,153.2)
\Line(17,133)(23,127)
\Line(17,127)(23,133)
\ArrowLine(40,130)(110,130)\Text(75,123)[lc]{{\scriptsize$\tilde{\chi}_j^0$}}
\ArrowLine(0,50)(40,50)\Text(20,57)[rc]{\scriptsize $\mu_L$}
\ArrowLine(150,50)(110,50)\Text(140,57)[lc]{\scriptsize $e_L^c$}
\Photon(115.4,99.42)(97,64.8){2}{7}\Text(122,105)[lt]{\scriptsize $\gamma$}
\DashCArc(75,29.8)(40.4,30,60){4}\Text(95,70)[rb]{{\scriptsize 
 $\tilde{e}_{R}$}}
\DashCArc(75,29.8)(40.4,60,90){4}
\DashCArc(75,29.8)(40.4,90,150){4}\Text(55,70)[rb]{{\scriptsize 
$\tilde{\mu}_{L}$}}
\Line(72,73.2)(78,67.2)
\Line(72,67.2)(78,73.2)
\Line(72,53)(78,47)
\Line(72,47)(78,53)
\ArrowLine(75,50)(40,50)\Text(75,40)[cc]{\scriptsize$\tilde{\chi}_j^0$}
\ArrowLine(75,50)(110,50)
\end{picture}
\end{center}
\vs{-1}

\noindent Figure 3. Top figure: one of the graphs contributing to $\taumg.$
Others (with $L\leftrightarrow R$, with a $\tilde\nu_{\tau}$-chargino loop, and
with the photon line attached to $\tilde\tau$) are not shown. Bottom figure:
similarly for $\mueg$.\vs{1}

\section{RESULTS}The results,
displayed in Fig.~(4),
are insensitive to $A_0$ and $sgn(\mu),$ and vary only slowly with $\tan\beta$
for this quantity in the range $3-10.$ The quantity $x$ referred to in the
caption to Fig.~(4) is a measure of the large renormalization of the
$\tilde\tau$ mass at GUT, $m_{\tilde\tau}^2|_{\rm GUT}=xm_{16}^2|.$ 
\begin{figure}[h]
\vspace{9pt}
\epsfig{figure=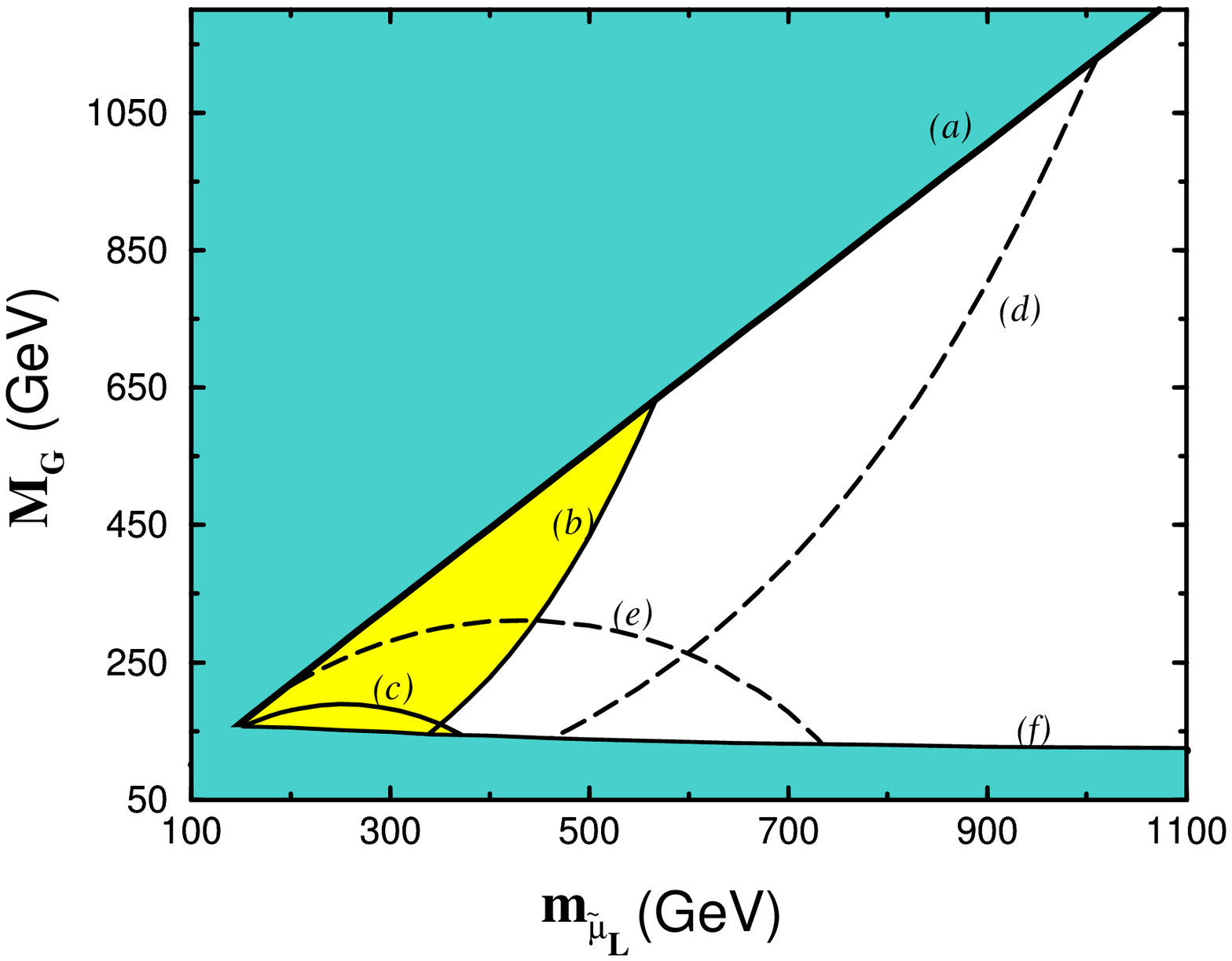,height=6truecm,width=7truecm}
\end{figure}

\noindent Figure 4. $(M_G, m_{\tilde\mu_L})$ parameter space excluded by
present and (possible) future data (all curves are for $\tan\beta=3, \mu>0,
A_0=1).$ Area below line (f) is excluded by direct chargino search at the
$M_Z$ pole. Area above line (a) is excluded by the R-G
analysis (see discussion in
text). Area between lines (a) and (b) is excluded
by present upper limit on $BR(\mu\rightarrow e\gamma)$.
Area between line (c) ($x=0.5$) and axes is  excluded
by present upper limit on
$\tau\rightarrow \mu\gamma$.
Lines (d) and (e) show the range of parameters
excluded if current limits were decreased by a factor of 10.\smallskip

The principal results (and conclusions) which can be gleaned from Fig.~(4) 
can be summarized as follows: 

\bi
\item  The breaking of
lepton-down quark universality at \mgut, as 
exemplified by the Georgi-Jarlskog
texture, is most naturally  implemented by imposing a Higgs structure 
with different
representation content in the 2-2 sector. 
\item This can induce substantial  non-universality in the $\tilde\mu-\tilde e$
sector as a result of  radiative corrections between \mpl\ and \mgut.  
\item In SO(10), this non-universality is greatly enhanced because of 
large group-theoretic factors. The result is manifested in a decay rate 
for $\mueg$ 
which impinges on present experimental limits for certain interesting regions
of MSSM parameter space. If current limits were decreased
by a factor of 10, a large portion of the slepton mass range would be excluded.
\item Conversely, either or both of $\mueg$ and $\taumg$ could be observed for
superpartner masses in the the few hundred GeV range at the price of this
factor of 10 improvement in statistics.
\item It should be stressed that physical  basis underlying the results
obtained here, namely a breaking of universality in the 
{\em representation content} of 
the 2-2 Higgs assignment,
will lead to a substantial rate for $\mueg$ in any other SUSY-GUT model.
\ei

\ed